\documentclass{aastex631}

\shorttitle{The Love number $k_{42}$}
\shortauthors{Idini \& Stevenson}
\graphicspath{{./}{figures/}}

\begin{document}

\title{The lost meaning of Jupiter's high--degree Love numbers}

\correspondingauthor{Benjamin Idini}
\email{bidiniza@caltech.edu}

\author[0000-0002-2697-3893]{Benjamin Idini}
\affiliation{Division of Geological and Planetary Sciences, California Institute of Technology \\
1200 E California Blvd, MC 150-21 \\
Pasadena, CA 91125, USA}

\author[0000-0001-9432-7159]{David J. Stevenson}
\affiliation{Division of Geological and Planetary Sciences, California Institute of Technology \\
1200 E California Blvd, MC 150-21 \\
Pasadena, CA 91125, USA}



\begin{abstract}
NASA's Juno mission recently reported Jupiter's high--degree (\added{degree $\ell$, azimuthal order $m$} $=4,2$) Love number $k_{42}=1.289\pm0.063$ \added{($1\sigma$), an order of magnitude above the hydrostatic $k_{42}$ obtained in a nonrotating Jupiter model. After numerically modeling rotation,} the hydrostatic $k_{42}=1.743\pm0.002$ is \added{still} $7\sigma$ away from the observation, \deleted{(i.e., equilibrium tides) numerically obtained by \cite{wahl2020equilibrium} in a Jupiter model that fits zonal gravity up to $J_4$} raising doubts about our understanding of Jupiter's tidal response. Here, we use first--order perturbation theory to explain the hydrostatic $k_{42}$ result analytically. We use a simple Jupiter equation of state ($n=1$ polytrope) to obtain the fractional change in $k_{42}$ when comparing a rotating model with a nonrotating model. Our analytical result shows that the hydrostatic $k_{42}$ is dominated by the tidal response at $\ell=m=2$ coupled into the spherical harmonic $\ell,m=4,2$ by the planet's oblate figure. The $\ell=4$ normalization in $k_{42}$ introduces an orbital factor $(a/s)^2$ into $k_{42}$, where $a$ is the satellite semimajor axis and $s$ is Jupiter's average radius. As a result, different Galilean satellites produce a different $k_{42}$. We conclude that high--degree tesseral Love numbers ($\ell> m$, $m\geq2$) are dominated by lower--degree Love numbers and thus provide little additional information about interior structure, at least when they are primarily hydrostatic. Our results entail important implications for a future interpretation of the currently observed \replaced{$7\sigma$ discrepancy in}{Juno} $k_{42}$. After including the coupling from the well--understood $\ell=2$ dynamical tides ($\Delta k_2 \approx -4\%$), Jupiter's hydrostatic $k_{42}$ requires an unknown dynamical effect to produce a fractional correction $\Delta k_{42}\approx-11\%$ in order to fit Juno's observation \added{within $3\sigma$. Future work is required to explain the required $\Delta k_{42}$.}
\end{abstract}



\section{Introduction} \label{sec:intro}

NASA's Juno mission recently registered Jupiter's tidal response in a set of Love numbers, observing \added{$k_{42}=1.289\pm0.063$ $(1\sigma)$ at the mid--mission perijove 17 \citep{durante2020jupiter}. The observation requires imposing the same Love number for all Galilean satellites as an a priori constraint, resulting in a Juno $k_{42}$ that represents the dominant tidal influence of Io \citep{durante2020jupiter}. Ignoring the effect of rotation on tides, the modeled hydrostatic tidal response is $k_{42}=0.1279$ in a Jupiter model with a density profile that fits the radius and zonal gravity up to $J_4$ \citep{wahl2020equilibrium}. In the same Jupiter model, numerical modeling indicates that rotation increases $k_{42}$ by an order of magnitude to $k_{42}=1.743\pm0.002$ in the case of Io's tidal forcing \citep{wahl2020equilibrium}, which is $7\sigma$ above the Juno $k_{42}$. The purpose of this paper is to illuminate on the rotational effect that leads to an order of magnitude enhancement in $k_{42}$, which is key to a correct interpretation of Juno's $k_{42}$.

} 

\deleted{an intriguing $7\sigma$ disagreement between the Juno $k_{42}$ and the hydrostatic $k_{42}$ calculated for a Jupiter model with a density profile that fits the radius and zonal gravity up to $J_4$ \citep{durante2020jupiter,wahl2020equilibrium}.} 


\added{Love numbers traditionally represent an interior property of the planet.} \added{A property of the planet must be independent of forcing; for example, the adiabatic sound speed.} The Love number $k_{42}$ corresponds to the tidal gravitational potential of Jupiter normalized by the tidal forcing produced by the satellite, both in the $\ell,m=4,2$ term of the spherical harmonics projection. The oblate figure of a rotating planet distorts the traditional meaning attributed to Love numbers by introducing spherical harmonic \replaced{mixing}{coupling}; \added{that is,} the tidal forcing at given $\ell$ produces a tidal response in multiple spherical harmonics. 
In particular, Jupiter's Love number $k_{42}$ contains a small contribution from the tidal response to the $\ell,m=4,2$ tidal forcing, plus a dominant contribution from the coupled tidal response to the $\ell=m=2$ tidal forcing. 
For the sake of brevity, we partially omit further references to the order $m$, which should always be considered $m=2$ throughout this paper.

The \added{coupled tidal response promoted by the} oblate figure of \replaced{a rotating planet}{Jupiter} enhances Jupiter's hydrostatic Love number $k_{42}$ by an order of magnitude when compared to a hypothetical nonrotating Jupiter \citep{wahl2020equilibrium}. To order of magnitude, we can estimate $k_{42}\sim qk_2(a/R)^2$, where $a$ is the semimajor axis of the satellite, $R$ is the planetary radius, $k_2$ is the $l=m=2$ Love number, and $q$ is the adimentional rotational parameter,
\begin{equation}
    q = \frac{\Omega^2s^3}{\mathcal{G}M}\textnormal{,}
\end{equation}
where $M$ is the mass of the planet, $\Omega$ is the planet's rotational frequency, $s$ is the average planetary radius, and $\mathcal{G}$ is the gravitational constant. In the case of Jupiter-Io, we obtain  $q\approx0.09$ and $k_{42}\sim1.9$. The coupled tidal response to the $\ell=2$  tidal forcing that contributes to $k_{42}$ is of order $\sim qk_2$. The factor $(a/R)^2$ describes how much smaller the tidal forcing is at $\ell=4$ when compared to $\ell=2$. \added{As numerically shown by \cite{wahl2020equilibrium},} the resulting hydrostatic Love number $k_{42}$ varies greatly among the Galilean satellites according to the semimajor axis of each orbit, a result that contradicts the traditional concept of the hydrostatic Love number defined as a property of the planet.  

Here, we use first--order perturbation theory to analytically explain the correction to the hydrostatic Love number introduced by rotation, a result only known so far via implementation of numerical strategies \citep{wahl2017concentric}. As a response to rotation, the oblate figure of the planet promotes mixing in the tidal response at different zonal degree $\ell$, causing a $\sim +10\%$ correction to $k_2$ observed both in Saturn \citep{lainey2017new,wahl2017concentric} and Jupiter \citep{durante2020jupiter,wahl2020equilibrium,idini2021dynamical}, and an order of magnitude increment in $k_{42}$ that is key to correctly \replaced{interpret}{interpreting} Juno's $k_{42}$ observation. 

The remainder of this paper is organized as follows. In Section 2, we derive the general solution for the hydrostatic tidal response in the interior of a gas giant planet mostly made of H-He. In Section 3, we use first--order perturbation theory to obtain the hydrostatic Love numbers while including the oblate figure of the planet introduced by rotation. In Section 4, we use our theoretical results from Section 3 to calculate Jupiter's hydrostatic $k_{42}$. In Section 5, we discuss the implications of our results. In Section 6, we summarize our conclusions.

\section{The hydrostatic tidal response}
In this section, we derive the equation and general solution for the hydrostatic tidal response of a gas giant planet mostly made of H-He fluid. Tides in hydrostatic equilibrium follow Poisson's equation and a simple equation of motion
\begin{equation}
    \nabla^2\phi=-4\pi\mathcal{G}\rho\textnormal{,}
    \label{eq:poisson}
\end{equation}
\begin{equation}
    \nabla p = \rho\nabla\phi\textnormal{.}
    \label{eq:momentum}
\end{equation}
The potential $\phi$ represents the relevant gravitational forces, $p$ is pressure, and $\rho$ is density.
In a gas giant planet mostly made of H-He, the equation of state of the fluid can be conveniently approximated by an $n=1$ polytrope \added{\citep{stevenson2020jupiter}}, which follows 
\begin{equation}
    p=K\rho^2\textnormal{,}
\end{equation}
where $K$ is a constant describing the material properties. The simple result $\nabla p =2K\rho\nabla\rho$ combined with Equations (\ref{eq:poisson}) and (\ref{eq:momentum}) allows us to obtain
\begin{equation}
    \frac{\nabla^2\phi}{k^2} + \phi =0\textnormal{,}
    \label{eq:helm}
\end{equation}
where $k^2=2\pi\mathcal{G}/K$.

The first approximation to the hydrostatic tidal response comes from considering tides as a perturbation $\phi'$ over a spherical planet with a spherically symmetric gravitational potential $\phi_0$. \added{Perturbation theory correctly approximates the tidal response because the tidal gravitational potential only constitutes a $\sim 10^{-6}$ part of the total gravitational potential.} In such scenario, the potential $\phi$ can be written as
\begin{equation}
    \phi = \phi_0 + \phi' + \phi_T\textnormal{,}
    \label{eq:perturbation}
\end{equation}
where the tidal forcing potential \replaced{assumes}{takes} the form
\begin{equation}
    \phi_T = \sum_{\ell=2}^\infty\sum_{m=-\ell}^\ell U_\ell^m\left(\frac{r}{s}\right)^\ell Y_\ell^m\textnormal{,}
    \label{eq:forcing}
\end{equation}
$U_{\ell,m}$ is a numerical factor defined by
\begin{equation}
    U_{\ell,m} = \left(\frac{s}{a}\right)^\ell\left(\frac{\mathcal{G}m_s}{a}\right) \left( \frac{4\pi (\ell-m)!}{(2\ell + 1)(\ell+m)!}\right)^{1/2} \mathcal{P}_l^{m}(0)\textnormal{,}
    \label{eq:Ulm}
\end{equation}
$\mathcal{P}_\ell^m$ are the associated Legendre polynomials of degree $\ell$ and azimuthal order $m$, $Y_\ell^m$ are spherical harmonics, and $m_s$ the mass of the satellite. The \added{hydrostatic} tidal response that solves Equation (\ref{eq:helm}) follows \added{\citep{idini2021dynamical}}
\begin{equation}
    \phi'= \sum_{\ell=2}^\infty\sum_{m=-\ell}^\ell \left(A_\ell j_\ell(kr)-\left(\frac{r}{s}\right)^\ell\right) U_\ell^mY_\ell^m\textnormal{,}
    \label{eq:phi_sph}
\end{equation}
where $j_\ell$ is the spherical Bessel function of the first kind.

 The boundary condition for $\phi'$ at the outer boundary of the planet defines the coefficients $A_\ell$. At the outer boundary of the planet, the tidal response $\phi'$ should match an external potential that decays with distance $r$ away from the planet, according to the factor $(s/r)^{\ell+1}$. Both potentials should also match in their directional derivative normal to the outer boundary of the planet. In a spherical planet, the directional derivative is simply $\partial_r$ due to the convenient decomposition of the tidal response into an axially symmetric factor and a spherical harmonic. In an oblate planet, however, the directional derivative involves additional terms that depend on the oblate figure of the planet. Additionally, the boundary of the planet is no longer the average radius $s$, but instead an oblate figure that varies with colatitute $\theta$ (i.e., roughly following a $\mathcal{P}_2(\cos\theta)$ figure). In the following section, we consider those effects to calculate $A_\ell^m$ and obtain first order corrections to the hydrostatic Love number due to the oblate figure of a rotating planet. 

\section{The hydrostatic tidal response in a rotating
planet\label{sec:pert}}

In this section, we use first--order perturbation theory to illuminate \added{on} the effect \replaced{of}{that} the oblate figure of a rotating planet \added{has} on the hydrostatic Love number. To $q$ order of approximation, the oblate figure of a gas giant mostly made of H-He follows (Appendix \ref{sec:oblatefig})
\begin{equation}
    R (\theta) \approx s\left(1-\frac{5}{\pi^2}q\mathcal{P}_2(\cos\theta)\right)\textnormal{.}
\end{equation}

At the outer boundary of the oblate planet (i.e., $r=R(\theta)$), the gravitational tidal response requires to satisfy the boundary condition
\begin{equation}
    \nabla\phi'(R)\cdot\hat{n} = \nabla \Theta'(R)\cdot\hat{n}\textnormal{,}
\end{equation}
where $\Theta$ is an external gravitational potential that matches $\phi'$ at $r=R$,
\begin{equation}
    \Theta_\ell^m{'} = \left(\frac{R}{r}\right)^{\ell+1}\phi_\ell^m{'}(R)\textnormal{,}
    \label{eq:ext}
\end{equation}
\begin{equation}
    \partial_r \Theta_\ell^m{'}= -\frac{(\ell+1)}{r}\left(\frac{R}{r}\right)^{\ell+1}\phi_\ell^m{'}(R)\textnormal{,}
\end{equation}
\begin{equation}
    \partial_\theta \Theta_\ell^m{'} = \frac{(\ell+1)}{R}\left(\frac{R}{r}\right)^{\ell+1}\phi_\ell^m{'}(R)\partial_\theta R+\left(\frac{R}{r}\right)^{\ell+1}\partial_\theta \phi_\ell^m{'}(R)\textnormal{,}
\end{equation}
and $\hat{n}$ is the vector normal to the oblate surface of the planet $R(\theta)$,
\begin{equation}
    \hat{n} = \left(1-\frac{5}{\pi^2}q\mathcal{P}_2(\cos\theta)\right)\hat{r} - \frac{15}{\pi^2}q\cos\theta\sin\theta\hat{\theta}\textnormal{.}
\end{equation}
After applying the differential operator in spherical coordinates and keeping only terms of order $q$, the boundary condition reduces to
\begin{equation}
    \sum_\ell\left(1-\frac{5}{\pi^2}q\mathcal{P}_2(\cos\theta)\right)\partial_r \phi_\ell^m{'}(R) = \sum_\ell-\frac{(\ell+1)}{s}\phi_\ell^m{'}(R)
    \label{eq:bc0}
\end{equation}

Assuming that rotation only causes a small deviation from a sphere, we can write $R = s(1-\epsilon \mathcal{P}_2)$, where $\epsilon = 5q/\pi^2$ is a small parameter. We evaluate the hydrostatic tidal response at the oblate outer boundary of the rotating planet by Taylor expansion of $\phi'$ over $\epsilon$,
\begin{equation}
    \phi_\ell^m{'}(R)\approx \phi_\ell^m{'}(s) - \epsilon s \partial_r\phi_\ell^m{'}(s)\mathcal{P}_2\textnormal{,}
    \label{eq:phi_oblate}
\end{equation}
\begin{equation}
    \partial_r \phi_\ell^m{'}(R)\approx \partial_r \phi_\ell^m{'}(s) - \epsilon s \partial_{r,r}\phi_\ell^m{'}(s)\mathcal{P}_2\textnormal{.}
    \label{eq:phip_oblate}
\end{equation}
From Equation (\ref{eq:phi_sph}), the hydrostatic tidal response of a sphere evaluated at $r=s$ follows
\begin{equation}
    \phi_\ell^m{'}(s) = (A_\ell j_\ell(ks)-1) U_\ell^m Y_\ell^m\textnormal{,}
\end{equation}
\begin{equation}
   \partial_r \phi_\ell^m{'}(s) = \left(A_\ell \partial_r j_\ell(ks)-\frac{\ell}{s}\right) U_\ell^m Y_\ell^m\textnormal{,}
\end{equation}
\begin{equation}
    \partial_{r,r}\phi_\ell^m{'}(s) =\left( A_\ell \partial_{r,r}j_\ell(ks)-\frac{\ell(\ell-1)}{s^2} \right) U_\ell^m Y_\ell^m\textnormal{.}
\end{equation}

We replace Equations (\ref{eq:phi_oblate}) and (\ref{eq:phip_oblate}) into Equation (\ref{eq:bc0}) to obtain the final equation for the coupled hydrostatic tidal response of an oblate rotating planet,
\begin{eqnarray}
    \sum_\ell \left( A_\ell \left(j_\ell(ks)\left(\frac{\ell+1}{s}\right)+\partial_r j_\ell(ks)\right) - \frac{2\ell+1}{s} \right) U_\ell^m Y_\ell^m \nonumber\\ = \frac{15}{\pi^2}q\sum_\ell \left( A_\ell \left( s\partial_{r,r}j_\ell(ks)+(\ell+2)\partial_r j_\ell(ks) \right) - \frac{\ell(2\ell+1)}{s}\right) U_\ell^m Y_\ell^m  \mathcal{P}_2\textnormal{.}
    \label{eq:final}
\end{eqnarray}

We use a recursive relation based on Clebsch-Gordan coefficients to calculate the coupling in spherical harmonics introduced by the term $ Y_\ell^m\mathcal{P}_2$ \citep{idini2021dynamical},
\begin{equation}
    Y_\ell^m\cos^2\theta = p_{\ell-1} p_\ell Y_{\ell-2}^m + (p_\ell^2 + p_{\ell+1}^2) Y_\ell^m + 
 p_{\ell+1} p_{\ell+2} Y_{\ell+2}^m\textnormal{,}
\end{equation} 
\begin{equation}
    p_\ell = \left(\frac{\ell^2 - m^2}{4\ell^2 -1}\right)^{1/2}\textnormal{.}
\end{equation}
Using the recursive relation above, we can write the term that couples the spherical harmonics of the hydrostatic tidal response as
\begin{equation}
    Y_\ell^m \mathcal{P}_2 = \frac{3}{2}\cos^2\theta Y_\ell^m - \frac{Y_\ell^m}{2} = \frac{3}{2}p_{\ell-1}p_\ell Y_{\ell-2}^m + \left(\frac{3}{2}(p_\ell^2+p_{\ell+1}^2)-\frac{1}{2}\right)Y_\ell^m + \frac{3}{2}p_{\ell+1}p_{\ell+2}Y_{\ell+2}^m\textnormal{.}
    \label{eq:coupled}
\end{equation}
In the following section, we \added{use Mathematica \citep{wolfram1999mathematica} to} evaluate Equations (\ref{eq:Ulm}), (\ref{eq:final}), and (\ref{eq:coupled}) to obtain $A_\ell$ in the case of Jupiter when tidally perturbed by the gravitational pull of Io.

\section{Jupiter's hydrostatic Love numbers}
 For the sake of simplicity, we analyze the case of coupling between Jupiter's rotational and hydrostatic tidal responses in $\ell=2$ and $\ell=4$, ignoring terms of higher degree. \added{To order of magnitude, the contribution to $k_{42}$ from $\ell=6$ tides follows $\sim qk_{62}(R/a)^2\sim q^2 k_{42}$, a second-order correction in $q$ and thus neglected here}. From Equation (\ref{eq:final}), we can write a linear system of equations in the form
 \begin{equation}
     c_1 A_2 + c_2 A_4 = c_3\textnormal{,}
     \label{eq:sys1}
 \end{equation}
 \begin{equation}
     c_4 A_2 + c_5 A_4 = c_6 \textnormal{,}
     \label{eq:sys2}
 \end{equation}
 where we obtain the $c_n$ coefficients from evaluating the sum in Equation (\ref{eq:final}) truncated at $\ell =4$ and set for $m=2$. 
 
  \begin{deluxetable*}{ccccccc}
\tablenum{1}
\tablecaption{Rotational correction to Jupiter's hydrostatic Love number under the tidal perturbation of Io, Europa, and Ganymede.\label{tab:love}}
\tablewidth{0pt}
\tablehead{
\colhead{}  & \colhead{polytrope} &\colhead{polytrope}&\colhead{polytrope} & \colhead{CMS}  & \colhead{CMS} & \colhead{CMS}\\
\colhead{} & \colhead{Io}& \colhead{Europa}& \colhead{Ganymede} & \colhead{Io} & \colhead{Europa}& \colhead{Ganymede}
}
\decimalcolnumbers
\startdata
$\delta k_2$     & 1.11  & 1.11& 1.11 &1.10&1.10 &1.10\\
$ \delta k_{42}$  & 14.4 & 35.0 & 87.6& 13.6 & 32.8 & 83.7\\
 \enddata
\tablecomments{The rotational correction $\delta k_{\ell m}$ is the ratio between the Love number in an oblate rotating Jupiter model, over the Love number in a spherical nonrotating Jupiter model. Jupiter's rotation rate follows $q = 0.0892$. (2-4) We obtain the analytical results in an $n=1$ polytrope from the ratio between Equation (\ref{eq:loveoblate}) and Equation (\ref{eq:lovesphere}). (5-7) We calculate \added{$\delta k$} from Love numbers reported in \cite{wahl2020equilibrium}, \added{which were numerically obtained with} the Concentric Maclaurin Spheroid (CMS) method. \deleted{numerical results}  }
\end{deluxetable*}
 
 Our analytical polytropic Jupiter model approximates the rotational correction to the hydrostatic Love number reported in \cite{wahl2020equilibrium} to $q$--order accuracy (Table \ref{tab:love}). In the case of tides raised by Io, we obtain a $\sim10\%$ increment in $k_2$ and an order of magnitude increment in $k_{42}$, both results previously reported in numerical calculations using the Concentric Maclaurin Spheroids (CMS) method \citep{wahl2016tidal,wahl2020equilibrium}. The order of magnitude enhancement in $k_{42}$ comes from the the tidal response to the $\ell=2$ tidal forcing rotationally--coupled into the $\ell=4$ gravitational field.
 We calculate the rotational correction $\delta k_{\ell m}$ in Table \ref{tab:love} as the ratio between the Love number in an oblate rotating polytrope over the Love number in a spherical nonrotating polytrope,
 \begin{equation}
     \delta k_{\ell m} = \frac{k_{\ell}}{k_\ell^{(s)}}\textnormal{.}
     \label{eq:dk}
 \end{equation}
 The Love number in a spherical nonrotating polytrope follows \citep{idini2021dynamical}
 \begin{equation}
     k_{\ell}^{(s)} = \left(\frac{2\ell +1}{\pi}\right) \frac{j_\ell(\pi)}{j_{\ell-1}(\pi)} - 1 \textnormal{,}
     \label{eq:lovesphere}
 \end{equation}
 while the Love number in an oblate rotating polytrope follows \added{(i.e., from evaluating Equations (\ref{eq:forcing}) and (\ref{eq:phi_sph}) at $r=s$)}
 \begin{equation}
     k_\ell = A_\ell j_\ell(\pi)-1 \textnormal{,}\label{eq:loveoblate}
 \end{equation}
 where the $A_\ell$ coefficients come from solving Equations (\ref{eq:sys1}) and (\ref{eq:sys2}). In the case of Jupiter's rotation ($q=0.0892$) and Io's semimajor axis, we obtain $A_2=5.19$ and $A_4=42.2$. For the sake of comparison, the nonrotating $n=1$ polytrope in hydrostatic equilibrium produces $A_2=5$ and $A_4=17.3$. 

We can apply our rotational corrections calculated from Equation (\ref{eq:dk}) to the nonrotating Love numbers of a Jupiter model with a more realistic equation of state and density profile (i.e., \cite{wahl2020equilibrium}). Our \added{Io} results agree with the CMS results within a margin of $2\%$ and $3\%$ for $k_2$ and $k_{42}$, respectively (Table \ref{tab:love2}). The difference between both results comes from second order effects not included in our analysis. \added{As we show here, the correct Love number in a rotating planet comes from the boundary condition that forces the smoothness of the tidal gravitational potential over an oblate planetary figure.}

\begin{deluxetable*}{cccccc}
\tablenum{2}
\tablecaption{Jupiter's hydrostatic Love numbers under the tidal perturbation of Io, Europa, and Ganymede.\label{tab:love2}}
\tablewidth{0pt}
\tablehead{
\colhead{}  & \colhead{CMS}&\colhead{$n=1$ polytrope} & \colhead{CMS} & \colhead{CMS} & \colhead{CMS}\\
\colhead{} & \colhead{Nonrotating} & \colhead{Io} & \colhead{Io} & \colhead{Europa}& \colhead{Ganymede}
}
\decimalcolnumbers
\startdata
$ k_2$     & 0.5364 & 0.60 & 0.5898 & 0.5894 & 0.5893\\
$ k_{42}$  &0.1279 & 1.8 & 1.7432 & 4.1975 & 10.7058\\
 \enddata
\tablecomments{(2) Numerical results obtained with the Concentric Maclaurin Spheroid (CMS) method applied to a nonrotating Jupiter model that follows an equation of state derived from ab initio simulations \citep{wahl2020equilibrium}. (3) We obtain the analytical results in an $n=1$ polytrope from applying the fractional correction in Equation (\ref{eq:dk}) to the nonrotating result in (2). (4-6) CMS Numerical results for the Love number of a rotating Jupiter model \citep{wahl2020equilibrium}. }
\end{deluxetable*}

\section{Discussion}
 
\subsection{The lost meaning of Love numbers in rotating gas giant planets\label{sec:meaning}}
High--degree tesseral Love numbers ($\ell >m$, $m\geq2$) lose their original meaning in a rotating planet with an oblate figure. As first proposed by A.E.H. Love in 1909, Love numbers represent the tidal response of a planet normalized by the tidal forcing, both at the same $\ell,m$ spherical harmonic. Accordingly, the hydrostatic tidal gravitational field of a spherical planet is a sum over terms $k_{\ell,m}\phi_{\ell,m}^T$. In this original meaning, the Love number represents an interior property of the planet. In the context of gas giant exoplanets, the Love number may describe the degree of central concentration of mass, with a lower Love number indicating a more centrally concentrated planet \citep{batygin2009determination}. However, the coupling introduced by rotation complicates this convenient picture. 

As shown in Equation (\ref{eq:phi_oblate}), the hydrostatic tidal response to the forcing at a given $\ell$ contains terms from multiple spherical harmonics. Rotation introduces a significant term with $\ell=4$ spherical harmonic corresponding to part of the tidal response to the $\ell=2$ tidal forcing. In fact, this term dominates Jupiter's $\ell=4$ tidal gravitational field, with $7\%$ of the amplitude arising from the tidal response to the $\ell=4$ tidal forcing and $93\%$ from the tidal response to the $\ell=2$ tidal forcing coupled by the oblate figure of the planet (Table \ref{tab:love}). According to this new term, the $\ell=4$ tidal gravitational field is (Equation (\ref{eq:phi_oblate})),
\begin{equation}
    \phi_4^2{'} \sim \frac{5}{\pi^2}q \phi_2^2{'}\mathcal{P}_2 \sim\frac{5}{\pi^2}q\left(\frac{r}{s}\right)^2k_2U_{2,2}Y_4^2 \textnormal{.}
    \label{eq:phi42}
\end{equation}

When computing $k_{42}$ after normalization of Equation (\ref{eq:phi42}) by the tidal forcing $\phi_4^2{_T}$, the term introduced by rotation promotes a dependency of $k_{42}$ on the semimajor axis of the satellite,
\begin{equation}
    k_{42}\sim \frac{5}{\pi^2}q\left(\frac{a}{s}\right)^2 k_2 \textnormal{.}
    \label{eq:k42}
\end{equation}
Explicitly revealed here using perturbation theory, the dependency of $k_{42}$ on semimajor axis was previously observed in numerical results obtained with CMS \citep{wahl2017concentric,wahl2020equilibrium}. The difference in $k_{42}$ among the Galilean satellites can be explained by the orbital factor $(a/s)^2$, where $a/s$ is roughly 6 for Io, 10 for Europa, and 15 for Ganymede (Table \ref{tab:love}). 

\subsection{The $7\sigma$ discrepancy observed by Juno\label{sec:juno}}



\deleted{Our analytical results add additional confidence to previous calculations of Jupiter's hydrostatic tidal response based on CMS \citep{wahl2020equilibrium}. Juno observations at the mid--mission perijove 17 indicate $k_{42}=1.289\pm0.063$ \added{$(1\sigma)$} \citep{durante2020jupiter}, which is $7\sigma$ away from the CMS hydrostatic result $k_{42}=1.743\pm0.002$ \citep{wahl2020equilibrium}.In a gas giant planet made mostly of H-He, we can use the $n=1$ polytrope to analytically obtain $k_{42}=0.12$ (Equation (\ref{eq:lovesphere})), which is not far from the nonrotating CMS result in a Jupiter model following an equation of state derived from ab initio simulations (Table \ref{tab:love2}).} 

Our \added{analytical} results \added{validate the accuracy of CMS to obtain the hydrostatic $k_{42}$ (Table \ref{tab:love2}),} \replaced{confirm}{confirming} a $7\sigma$ discrepancy between Juno's $k_{42}$ and hydrostatic tides \added{at perijove 17}. \deleted{ validating the accuracy of CMS to obtain the right hydrostatic $k_{42}$. } The CMS hydrostatic $k_{42}$ requires a $\Delta k_{42}\approx-15\%$ fractional correction to fit the Juno observation within $3\sigma$. Due to rotational coupling (Equation (\ref{eq:k42})), one part of the required correction comes from dynamical effects on $k_2$ that include the Coriolis effect. The $\Delta k_2\approx-4\%$ effect introduced by dynamical tides \citep{idini2021dynamical,Lai_2021} reduces the fractional correction required by Juno to $\Delta k_{42}\approx-11\%$. This residual effect must come from additional dynamical effects related to the $\ell=4$ tidal response, which is only a small fraction ($7\%$) of the hydrostatic $k_{42}$. At perijove 17, Juno $3\sigma$ uncertainty on $k_2$ is only $3\%$ \citep{idini2021dynamical}, much smaller than the required $\Delta k_{42}\approx-11\%$. Consequently, additional hypothetical dynamical effects applied to $k_2$ are constrained by Juno to be small and insufficient. We require future studies to understand the origin of the $\Delta k_{42}$ fractional correction required to fit Juno observations.

\added{ The uncertainty $\sigma=0.063$ at perijove 17 depends on imposing the same Love number for all Galilean satellites. When $k_{42}$ is let to freely vary  among satellites, an orbital resonance between the Juno spacecraft and Io--Europa--Ganymede (in mean--motion resonance 1:2:4) conspires against a unique decomposition of the joint tidal gravitational field, leading to a tradeoff among individual contributions that sharply increases uncertainty ($\sigma=0.353$ for Io). Future perijove passes from Juno's extended mission will break the tradeoff given a recent change in Juno's orbital period. Currently, the best representation of Jupiter's $k_{42}$ due to Io's gravitational pull comes from assuming a reasonable a priori constraint to the $k_{42}$ caused by the other satellites. Imposing the same Love number to all Galilean satellites equals to assume that Io dominates the $k_{42}$ tidal gravitational field, which is true unless Europa or Ganymede cause a tidal resonance with Jupiter. }

\section{Conclusions}
We used first--order perturbation theory to calculate the rotational correction to Jupiter's hydrostatic Love number $k_{42}$. We showed that the oblate figure of the rotating planet forces the $\ell=m=2$ tidal response to couple into the $\ell,m=4,2$ tidal gravitational field, increasing the \replaced{hydrostic}{hydrostatic} $k_{42}$ beyond an order of magnitude for Io and roughly by two orders of magnitude for Ganymede. As a result, we conclude that low--degree hydrostatic Love numbers dominate high--degree hydrostatic tesseral Love numbers ($\ell>m$, $m\geq2$), and thus the latter provide little additional information about interior structure. The exception is the case where dynamical effects particular to a given high--degree Love number acquire relevant amplitude due to, for example, tidal resonances.

Our analysis leads to important implications for the correct interpretation of a $7\sigma$ anomaly in Jupiter's $k_{42}$ as observed by NASA's Juno mission. The Juno $k_{42}$ anomaly is slightly attenuated by the coupled $\ell=2$ dynamical tides ($\Delta k_2\approx-4\%$). At Juno's mid--mision perijove 17, Jupiter's hydrostatic Love number $k_{42}$ requires an additional fractional correction $\Delta k_{42}\approx-11\%$ from unknown dynamical effects associated to its tidal response to the $\ell=4$ tidal forcing. We require further analysis to unravel the origin of the required fractional correction.


\begin{acknowledgments}
We acknowledge the support of NASA's Juno mission. We benefited from constructive discussions with James Fuller, Janosz Dewberry, and Christopher Mankovich.
\end{acknowledgments}

%

\vspace{5mm}


\software{ Mathematica \citep{wolfram1999mathematica}
          }


\appendix

\section{The oblate figure of a gas giant planet \label{sec:oblatefig}}
In this appendix, we revise the classical result of calculating the first--order figure of a gas giant planet mostly made of H-He (i.e., an $n=1$ polytrope) after \added{being} perturbed by the centrifugal effect \added{\citep{hubbard1984planetary}}. The external potential of a body perturbed by the centrifugal effect follows
\begin{equation}
    \phi \approx \frac{\mathcal{G}M}{r}\left(1-\left(\frac{s}{r}\right)^2J_2\mathcal{P}_2(\cos\theta)\right)\textnormal{,}
    \label{eq:extpot}
\end{equation}
and the rotational forcing potential follows
\begin{equation}
    \phi_R = \frac{\Omega^2r^2}{3}\left(1-\mathcal{P}_2(\cos\theta)\right)\textnormal{,}
\end{equation}
\added{where $\mathcal{G}$ is the gravitational constant, $M$ is the mass of the planet, $s$ is the average planetary radius, $\Omega$ is the planet's rotational frequency, $J_2$ is the zonal gravitational coefficient of degree $\ell=2$, and $\mathcal{P}_2$ the Legendre Polynomial of degree $\ell=2$.} The outer boundary of the planet represents an equipotential surface where pressure is constant. If $R(\theta)$ represents the outer boundary of the planet, we require to satisfy
\begin{equation}
    \phi(R)+\phi_R(R) = \textnormal{constant} = \phi(\theta=0)\textnormal{,}
\end{equation}
or
\begin{equation}
    \frac{\mathcal{G}M}{R}\left(1-\left(\frac{s}{R}\right)^2J_2\mathcal{P}_2(\cos\theta)\right) + \frac{\Omega^2R^2}{3}\left(1-\mathcal{P}_2(\cos\theta)\right) = \frac{\mathcal{G}M}{b}\left(1-\left(\frac{s}{b}\right)^2J_2\right)\textnormal{,}
    \label{eq:equipot}
\end{equation}
where $b$ is the polar radius. We consider the first--order expansion on the oblate figure of the planet as 
\begin{equation}
    R(\theta) = \sum_{\ell=0}^\infty \delta r_{2\ell}\mathcal{P}_{2\ell}(\cos\theta)\approx s +\delta r_2\mathcal{P}_2(\cos\theta)\textnormal{,}
\end{equation}
\added{where $\delta r_2$ is a $\ell=2$ perturbation to the figure of the planet.} After replacing the first--order expansion of $R(\theta)$ into Equation (\ref{eq:equipot}), we obtain the classical result
\begin{equation}
    \frac{\delta r_2}{s}\approx - \left(J_2 + \frac{q}{3}\right)\textnormal{,}
\end{equation}
which is accurate to first order in $q$, \added{and $q$ is the adimentional rotational parameter.}

To obtain $J_2$ in an $n=1$ polytrope, we \added{equal the external potential in Equation (\ref{eq:extpot}) to }the  $\ell=2$ rotational gravitational response of the polytrope to the centrifugal effect, both evaluated at $r=s$,
\begin{equation}
   J_2 = \left(5j_2(\pi)-1\right)\frac{q}{3}\textnormal{,}
\end{equation}
\added{where $j_2$ is the spherical Bessel function of the first kind. The rotational gravitational response in a $n=1$ polytrope follows the same equations and boundary conditions than the tidal gravitational response with $\phi_T$ replaced by $\phi_R$ in Equation (\ref{eq:perturbation}).} Finally, we obtain the first--order oblate figure of a rotating gas giant planet perturbed by the centrifugal effect,
\begin{equation}
    R (\theta) \approx s\left(1-\frac{5}{\pi^2}q\mathcal{P}_2(\cos\theta)\right)\textnormal{.}
\end{equation}


\bibliography{sample631}{}

\begin{thebibliography}{}
\expandafter\ifx\csname natexlab\endcsname\relax\def\natexlab#1{#1}\fi
\providecommand{\url}[1]{\href{#1}{#1}}
\providecommand{\dodoi}[1]{doi:~\href{http://doi.org/#1}{\nolinkurl{#1}}}
\providecommand{\doeprint}[1]{\href{http://ascl.net/#1}{\nolinkurl{http://ascl.net/#1}}}
\providecommand{\doarXiv}[1]{\href{https://arxiv.org/abs/#1}{\nolinkurl{https://arxiv.org/abs/#1}}}

\bibitem[{Batygin {et~al.}(2009)Batygin, Bodenheimer, \&
  Laughlin}]{batygin2009determination}
Batygin, K., Bodenheimer, P., \& Laughlin, G. 2009, The Astrophysical Journal
  Letters, 704, L49

\bibitem[{Durante {et~al.}(2020)Durante, Parisi, Serra, Zannoni, Notaro,
  Racioppa, Buccino, Lari, Gomez~Casajus, Iess, {et~al.}}]{durante2020jupiter}
Durante, D., Parisi, M., Serra, D., {et~al.} 2020, Geophysical Research
  Letters, 47, e2019GL086572

\bibitem[{Hubbard(1984)}]{hubbard1984planetary}
Hubbard, W.~B. 1984, Planetary interiors (Van Nostrand Reinhold)

\bibitem[{Idini \& Stevenson(2021)}]{idini2021dynamical}
Idini, B., \& Stevenson, D.~J. 2021, The Planetary Science Journal, 2, 69

\bibitem[{Lai(2021)}]{Lai_2021}
Lai, D. 2021, The Planetary Science Journal, 2, 122, \dodoi{10.3847/psj/ac013b}

\bibitem[{Lainey {et~al.}(2017)Lainey, Jacobson, Tajeddine, Cooper, Murray,
  Robert, Tobie, Guillot, Mathis, Remus, {et~al.}}]{lainey2017new}
Lainey, V., Jacobson, R.~A., Tajeddine, R., {et~al.} 2017, Icarus, 281, 286

\bibitem[{Stevenson(2020)}]{stevenson2020jupiter}
Stevenson, D.~J. 2020, Annual Review of Earth and Planetary Sciences, 48

\bibitem[{Wahl {et~al.}(2016)Wahl, Hubbard, \& Militzer}]{wahl2016tidal}
Wahl, S.~M., Hubbard, W.~B., \& Militzer, B. 2016, The Astrophysical Journal,
  831, 14

\bibitem[{Wahl {et~al.}(2017)Wahl, Hubbard, \& Militzer}]{wahl2017concentric}
---. 2017, Icarus, 282, 183

\bibitem[{Wahl {et~al.}(2020)Wahl, Parisi, Folkner, Hubbard, \&
  Militzer}]{wahl2020equilibrium}
Wahl, S.~M., Parisi, M., Folkner, W.~M., Hubbard, W.~B., \& Militzer, B. 2020,
  The Astrophysical Journal, 891, 42

\bibitem[{Wolfram(1999)}]{wolfram1999mathematica}
Wolfram, S. 1999, Assembly Automation

\end{thebibliography}
\bibliographystyle{aasjournal}



\end{document}